# Hierarchical Bayesian modeling of agrivoltaic dairy-calves under missing information

Fabiana A Demeu[1*], Abdias A Oliveira[2], Uziel Bemvindo[1], Ana K Salman[3], Jomel F Santos[1], Alex SC Maia[4], Thatiane AT Souza[5], Hugo FM Milan[5]

[1]Rondônia Technology Institute (IFRO), Ariquemes Campus

[2]São Francisco Farm, Ariquemes city, Rondônia state

[3]Brazilian Agricultural Research Corporation in Rondônia State (Embrapa)

[4]São Paulo State University (UNESP), Jaboticabal Campus

[5]Rondônia University (UNIR)

Corresponding Author: Fabiana A Demeu, email: fabiana.alves@ifro.edu.br

**Abstract**

Livestock 5.0 provides a framework to optimize animal welfare and productivity amidst escalating climate change and current pressures to sustainably increase production through limited expansion of agricultural lands, while equilibrating the food-water-energy nexus. Directly tackling the food-water-energy nexus, agrivoltaics emerges as a promising solution for dual-land use for food and renewable energy production. In this paper, we evaluated the effectiveness of agrivoltaics for dairy-calves in a commercial farm through a joint hierarchical Bayesian model, developed to account for non-linear and time-varying relationships, pedigree structure, and missing information. Uncertainty from missing information was propagated to the posterior estimates. While weight at birth and calf sex were not associated with serum Brix, we found that serum Brix was positively associated with colostrum Brix, which allowed to estimate the probabilities of the serum Brix quality (poor, fair, good, or excellent) given colostrum Brix. Poor serum Brix quality was positively associated with disease event rate, with similar rates for the other quality classes. Serum Brix quality was not associated with calf weight, indicating that farm management overcame the negative effects of low serum Brix quality. Previous number of diseases was positively associated with disease event rate, while weight at birth was negatively associated with disease event rate. Agrivoltaics were not associated with disease event rate, indicating that agrivoltaics can be implemented as a safe, dual-use system for increased calf comfort an welfare.

## Graphical abstract

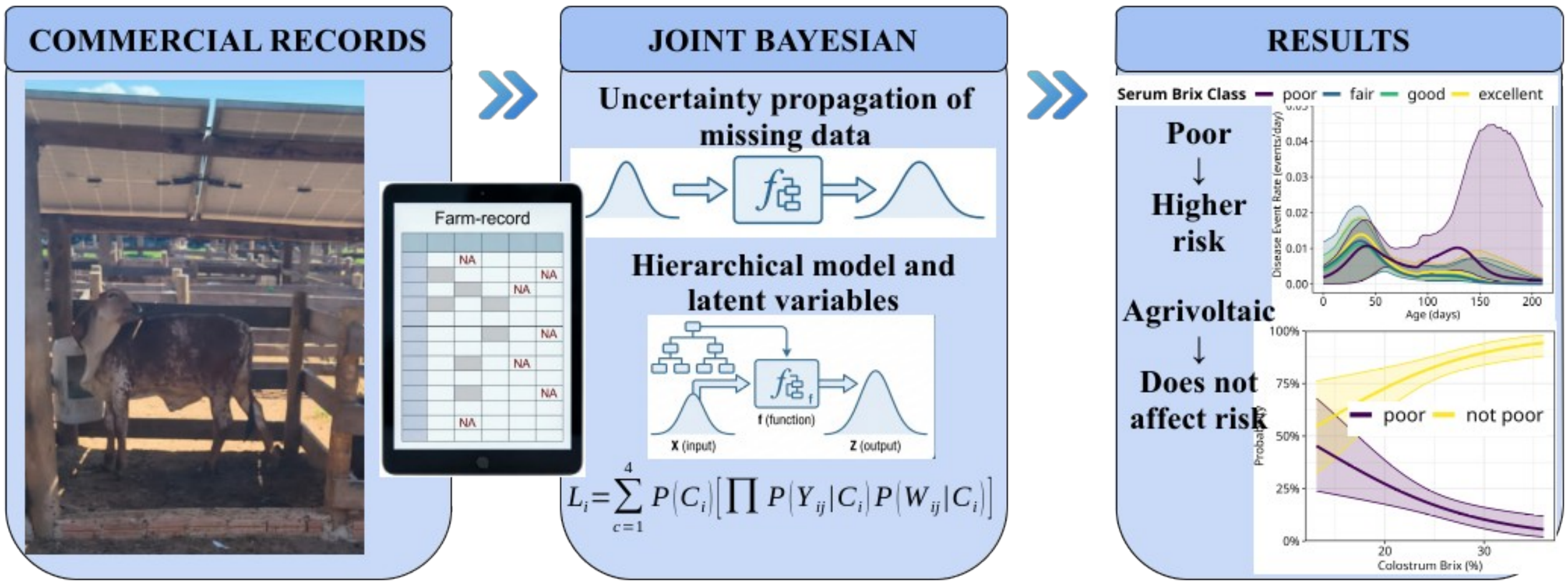

**Highlights**

- Joint hierarchical Bayesian model propagated uncertainty from missing data
- Serum Brix increased with colostrum Brix but not with birth weight nor sex
- Disease rate increased with poor serum Brix but not with other Brix qualities
- Weight trajectories showed no clear association with serum Brix quality
- Agrivoltaics did not increase disease rate, indicating to be a safe, dual-use system

## 1. Introduction

Society is entering the fifth industrial revolution (Industry 5.0) by emphasizing sustainability, resilience, and human-centric innovation alongside technological advancement (European Commission, 2021). Following the fifth industrial revolution, comes the fifth livestock revolution (Livestock 5.0; Fountas et al., 2024). In this new era, innovation is driven by the integration of digital technologies, automation, and data-driven decision-making to improve resource-use efficiency and animal welfare and comfort. Considering the importance of the livestock industry worldwide, it is paramount to develop Livestock 5.0 technologies to sustainably and comfortably manage animals under climate change.

Climate change, allied with non-sustainable farming practices, is imposing strong constraints into food security and energy production (Karri et al., 2026). Considering the complex environmental interplay with animal production, traditional farming practices were proven inefficient to solve the interconnected food-water-energy nexus problems (Gathala et al., 2020). Animal agrivoltaics has emerged as a state-of-the-art solution (Chopdar et al., 2024; Mamun et al., 2022) to improve animal welfare and comfort while avoiding land-use competition between food and energy production. Previous research has proven the viability of animal agrivoltaics and its benefits over conventional structures, resulting in a positive symbiosis that improves animal welfare and production (Faria et al., 2023; Maia et al., 2020) while reducing land-use competition (Fidelis et al., 2026; Fonsêca et al., 2023). The realization of the potential of agrivoltaics within Livestock 5.0 requires proper data management and analytical methods, especially in commercial farms that typically collect routine operational data rather controlled protocols, resulting in incomplete, irregular measurements. Currently, a gap persists in the literature regarding the validation of Livestock 5.0 methodologies to handle missing data and reliably assess agrivoltaics benefits.

To address this gap, we developed a joint hierarchical Bayesian model to characterize dairy calf health and weight trajectories using real, incomplete operational records from a commercial agrivoltaic farm, propagating uncertainty associated with missing data to the posterior estimates. Specifically, we evaluated the associations of 1) colostrum Brix, birth weight, and sex with serum Brix, 2) serum Brix quality class, previous disease count, birth weight, sex, and agrivoltaic exposure during the pre- and post-weaning periods with disease event rates, and 3) characterized longitudinal body-weight trajectories using the Brody growth equation. Using this unified Bayesian framework, we will assess wether agrivoltaics increase disease risks of dairy-calves in a commercial setting, contributing to the evaluation of data-driven approaches for Livestock 5.0.

## 2. Materials and Methods

### *2.1. Animal handling and data collection*

Data was obtained from records of the São Francisco Farm commercial dairy farm (Ariquemes city, Brazil, climate Aw, Alvares et al., 2014) from Giroland (Gir x Holstein) calves, with reproduction methods including artificial, embryo transfer, and natural. Calves were weighted after birth, given colostrum (10% live weight; as fast as possible), and navel treated (10% iodine). If colostrum Brix was lower than 25%, colostrum was enriched with powdered colostrum or substituted with high Brix colostrum from the farm's colostrum bank. Calves were then managed in individual elevated pens, followed by individual agrivoltaic pens (~25 days-old; Fig. 1) where calves were weaned (~100 days-old). After weaning, animals were moved to collective pasture paddocks.

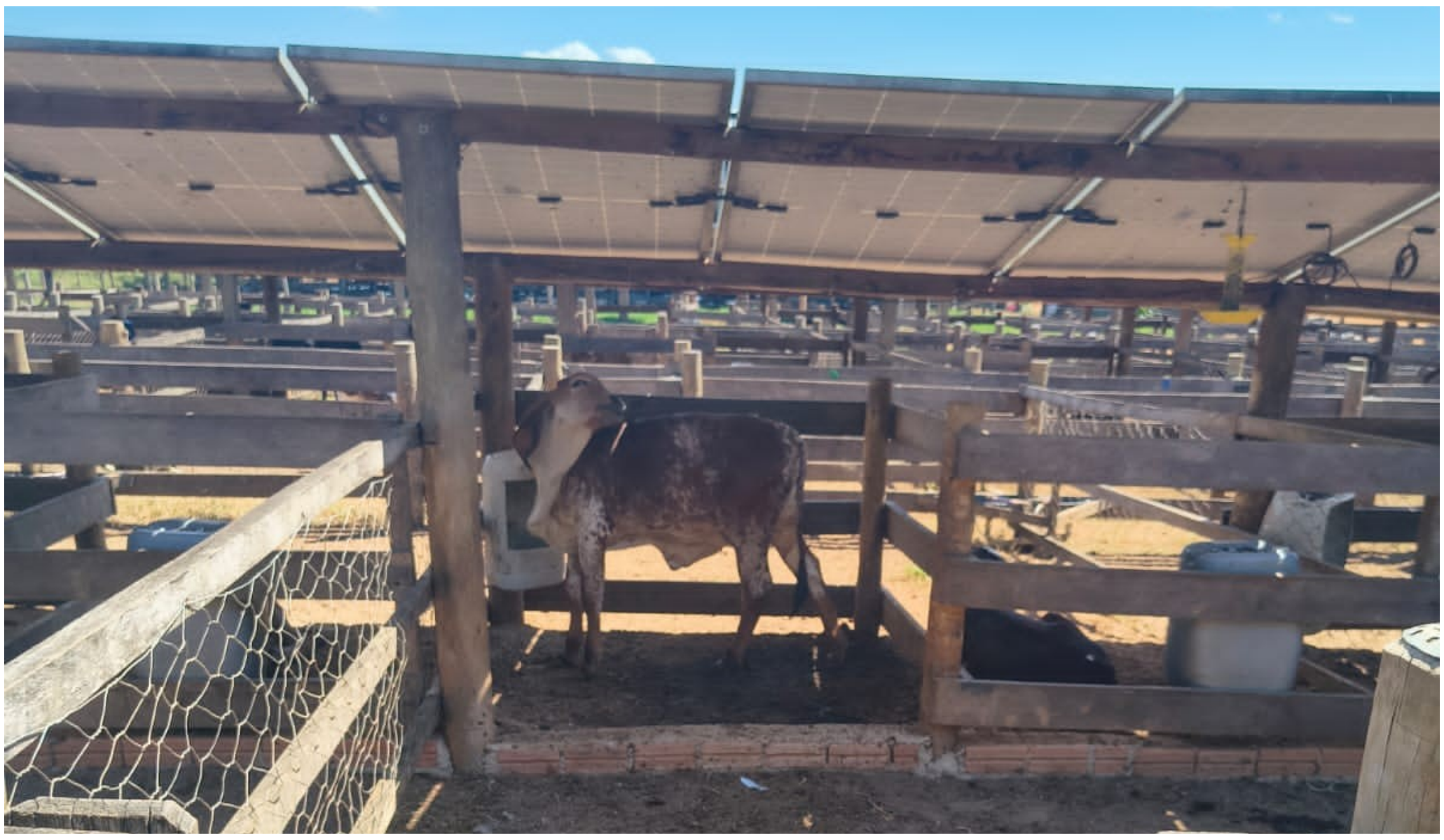

Figure 1. Individual agrivoltaic pen used as shading resources to improve dairy-calves comfort and welfare.

### *2.2. Bayesian Statistical Modeling*

Bayesian statistical analyses (Gelman et al., 2013; McElreath, 2020) were performed in the R/R-Studio (Posit Team, 2026; R Core Team, 2026) environment using the brms package (Bürkner, 2017, 2018, 2021). Priors were chosen based on preliminary tests (Milan, 2026) and to not restrict possible outcomes (Gelman et al., 2026). We assumed that weight at birth and colostrum serum affected Brix serum, which in turn affect immunity level, which in turn affect weigh gain of calves, modeled as a t-distribution to allow for outliers:

$$B_i \sim \beta_{B0} + \beta_{B1}\mathrm{CB}_i + \beta_{B2}\mathrm{WB}_i + \beta_{B3}\mathrm{Sex}_i + \mathrm{t}(\nu_B, \sigma_B) \tag{1}$$

$$\beta_{B0}, \beta_{B1}, \beta_{B2} \sim \mathrm{N}(0, 0.5) \tag{2}$$

$$\nu_B \sim \mathrm{Gamma}(2, 1) \tag{3}$$

$$\sigma_B \sim \mathrm{Gamma}(5, 0.5) \tag{4}$$

where,standardized = values subtracted by mean and divided by standard deviation, $B_i$ = standardized Brix serum level of calf *i*, $\mathrm{CB}_i$ = standardized colostrum Brix level of calf *i*, $\mathrm{WB}_i$ = standardized weight at birth of calf *i*, $\mathrm{Sex}_i$ = sex of calf *i*, $\beta$ = linear coefficients, $a \sim f(.)$ = *a* follows the statistical distribution $f(.)$, t(*a*, *b*) = t-student probability distribution with degrees of freedom *a* and standard deviation *b*, N(*a*, *b*) = normal probability distribution with mean *a* and standard deviation *b*, and Gamma(*a*, *b*) = gamma probability distribution with shape parameter *a* and rate parameter *b*.

Disease events were modeled as dependent on the Brix serum class (poor < 8.1%, 8.1% ≤ fair < 8.9%, 8.9% ≤ good < 9.4%, excellent ≥ 9.4%; Lombard et al., 2020) following a Poisson regression:

$$Y_{ij} | C_i = c \sim \mathrm{Poisson}(\lambda_{ij,c}) \tag{5}$$

$$\ln(\lambda_{ij,c}) = \mathrm{Base}_{ij} + \mathrm{Class}_c(t_{ij}) \tag{6}$$

$$\mathrm{Base}_{ij} = \ln(\mathrm{E}_{ij}) + \beta_{H0,e} + s_e(t_{ij}) + \beta_{H1}\mathrm{Mng}_{ij} + \beta_{H2}\mathrm{WB}_i + \beta_{H3}\mathrm{Sex}_i + \beta_{H4}\mathrm{PD}_{ij} + \alpha_{H,i} \tag{7}$$

$$\mathrm{Class}_c(t_{ij}) = \beta_{H0,c} + s_c(t_{ij}) \tag{8}$$

$$\alpha_{H,i} \sim \mathrm{N}(0, A\sigma^2_{H,\alpha}) \tag{9}$$

$$\beta_{H0,e} \sim \mathrm{N}(-6.2, 0.5) \tag{10}$$

$$\beta_{H1}, \beta_{H2}, \beta_{H3} \sim \mathrm{N}(0, 0.5) \tag{11}$$

$$\beta_{H4} \sim \mathrm{N}(0, 0.12) \tag{12}$$

$$\beta_{H0,p} \sim \mathrm{N}(0.5, 0.2) \tag{13}$$

$$\beta_{H0,f}, \beta_{H0,g} \sim \mathrm{N}(-0.2, 0.2) \tag{14}$$

$$\sigma_{H,\alpha} \sim \mathrm{Exp}(4) \tag{15}$$

$$b_{s,p} \sim \mathrm{Exp}(4) \tag{16}$$

$$b_{s,f}, b_{s,g} \sim \mathrm{Exp}(2) \tag{17}$$

$$b_{s,e} \sim \mathrm{t}(3, 2.5) \tag{18}$$

where, $Y_{ij}$ = disease event of calf *i* at time interval *j* (time intervals were defined as at most 5 days, being shorter when an event happened during that time interval, such as sickness or moving to agrivoltaics installation; e.g., if an animal became sick at day 3, the time intervals would be 1-3, 3-8, and so on), $C_i$ = Brix serum quality class *c* (*p* = poor, *f* = fair, *g* = good, *e* = excellent) of calf *i*, $\lambda_{ij,c}$ = rate of disease event of calf *i* at time interval *j* with Brix serum quality class *c*, $\mathrm{Base}_{ij}$ = common base function for all *c*, $E_{ij}$ = length of the exposure time ($\Delta t$, in days; for instance, for 1-3 = 3; for 3-8 = 5) for time interval *j* and calf *i*, $s_c(t_{ij})$ = smooth spline learned for time *t* of time interval *j* of calf *i* and class *c* (smooth spline functions for class *c* are shared across all animals within class *c*; spline dimensions were defined as small as possible to not restrict the functions learned: poor = 3, fair = 3, good = 7, and excellent = 10; note that splines for poor, fair, and good represent deviations from the excellent class, chosen to be the base spline function due to its high number of samples when compared to the others), $\mathrm{Mng}_{ij}$ = management class (elevated pen: installation of newborn calves; agrivoltaics: next installation for older calves; agrivoltaics postweaning: weaned calves in the agrivoltaics installation) for calf *i* at time interval *j*, $\mathrm{PD}_{ij}$ = number of previous disease events for calf *i* at time interval *j*, $b_{s,c}$ = parameters for the smoothing splines for class *c*, $\alpha_i$ = random effect for the pedigree term (isolates the effects of the environment from animal genetics) for calf $i$, $A$ = parenting matrix (rows and columns represent calves id, with 1 when row = column—representing the same calf—, 1 when calves are brothers, 0.5 when half-brothers, and 0 otherwise; Mrode, 2014), $\sigma_a$ = standard deviation of the additive genetic characteristics, $a \mid b$ = $a$ conditional on $b$, Poisson($a$) = Poisson probability distribution function with expectation number of events $a$, and Exp($a$) = exponential probability distribution function with rate parameter $a$.

Animal weight was modeled following a non-linear Brody function (Brody, 1945), following a t-distribution to allow for outliers, parametrized for weight at birth:

$$W_{ik} | C_i = c \sim \exp(lW_i) + \exp(lG_i)\left[1 - \exp\left(-\exp(lK_i) t_{ik}\right)\right] + \mathrm{t}(v_w, \sigma_w) \quad (19)$$

$$lW_i = \beta_{W0} + \beta_{W1}\mathrm{Sex}_i + \alpha_{W,i} \quad (20)$$

$$lG_i = \beta_{G0,e} + \beta_{G1}\mathrm{Sex}_i + \beta_{G3,c \neq e} C_i + \alpha_{G,i} \quad (21)$$

$$lK_i = \beta_{K0,e} + \beta_{K1}\mathrm{Sex}_i + \beta_{K3,c \neq e} C_i \quad (22)$$

$$\beta_{W0} \sim \mathrm{N}(3.4, 0.2) \quad (23)$$

$$\beta_{W1}, \beta_{G1}, \beta_{K1} \sim \mathrm{N}(0, 0.1) \quad (24)$$

$$\beta_{G0,e}, -\beta_{K0,e} \sim \mathrm{N}(5.3, 0.2) \quad (25)$$

$$-\beta_{G3,c=p}, \beta_{G3,c=f}, \beta_{G3,c=g}, -\beta_{K3,c=p}, \beta_{K3,c=f}, \beta_{K3,c=g} \sim \mathrm{N}(-0.1, 0.2) \quad (26)$$

$$\sigma_w \sim \mathrm{Exp}(0.5) \quad (27)$$

$$v_W \sim \mathrm{Gamma}(2, 1) \quad (28)$$

$$\alpha_{W,i} \sim \mathrm{N}(0, A\sigma^2_{W,\alpha}) \quad (29)$$

$$\alpha_{G,i} \sim \mathrm{N}(0, A\sigma^2_{G,\alpha}) \quad (30)$$

$$\sigma_{W,i}, \sigma_{G,i} \sim \mathrm{Exp}(10) \quad (31)$$

where $W_{ik}$ = body weight of calf $i$ at time $k$, $lW_i$ = natural log of weight at birth for calf $i$, $lG_i$ = natural log of adjusting factor $G$ for calf $i$, $lK_i$ = natural log of adjusting factor $K$ for calf $i$.

### *2.3. Missing Data Modeling*

A major challenge working with commercial data is missing data (*e.g.*, serum and/or colostrum Brix data, and transition days between pens). Traditional approaches, such as listwise deletion, may introduce selection bias and reduce statistical power, as missigness in commercial settings is rarely completely at random (McElreath, 2020). To overcome this limitation, we employed a dual Bayesian strategy to propagate the uncertainty of missing information directly into the posterior estimates. For missing colostrum Brix, days in the elevated pen (which marked the transition to agrivoltaic pens), and weaning age, we utilized Multiple Imputation by Chained Equations (MICE; van Buuren and Groothuis-Oudshoorn, 2011), with the number of imputed datasets was set equal to the highest percentage of missingness (Bodner, 2008; von Hippel, 2020; White et al., 2011). The hierarchical Bayesian model (Eqs. 1-31) was fitted for all imputed datasets, with posterior draws from models fitted to each individual imputed datasets to capture the uncertainty of the imputation process.

As the Brix serum affects the hierarchical Bayesian model, its missing values were treated as latent states, estimated with the likelihood ($L_i$) calculated as a finite mixture model (Gelman et al., 2013), considering the weighted probability of calf $i$ belonging to class $c$ using the observed likelihood for health and body weight:

$$L_i = \sum_{c=1}^{4} P(C_i = c)\left[\prod P(Y_{ij} | C_i = c) P(W_{ij} | C_i = c)\right] \quad (32)$$

## 3. Results and Discussion

The dataset contained 392 disease records from 347 calves (233 females; 114 males) and, reflecting the reality of commercial farm operations, contained missing information: 65 (18.7%) calves without colostrum Brix, 150 (43.2%) calves without weaning age, 170 (49.0%) calves without days in elevated pen (which marked the transition to agrivoltaic pens), and 195 (56.2%) calves without serum Brix. In traditional frequentist and artificial intelligence frameworks, which rely on complete-case analysis (Efron and Hastie, 2016), using this dataset would result in eliminating 219 (63.1%) calves from the analysis, reducing statistical power and introducing selection bias (as missigness in commercial settings is rarely completely at random; McElreath, 2020). Instead, by using a joint hierarchical Bayesian approach (Eqs. 1-32), treating missing values as latent parameters and marginalizing unobserved data, the model propagated the uncertainty of the missing data to the posterior estimates (Gelman et al., 2026). Through uncertainty propagation, our inferences are statistically robust, unbiased, and a representative of the commercial heard dynamics (McElreath, 2020), highlighting the value of advanced analytics in Livestock 5.0 applications.

Colostrum Brix (Fig. 2a; 0.20: 0.08, 0.32; where *a*: *b*, *c*: *a* = posterior mean and *b*, *c* = lower, upper 95% credible interval; substantial effects do not cross 0.0) was positively associated with serum Brix, indicating passive immunity transfer between colostrum and serum (Lichtmannsperger et al., 2023) as Brix is associated with higher immunoglobulin G (IgG) concentrations (Sockett et al., 2023; Soufleri et al., 2023). Weight at birth (Fig. 2b; 0.12: -0.03, 0.26) was not associated with serum Brix, indicating that furnishing adequate colostrum volumes mitigate previously found negative correlations (Bandlow et al., 2023). Calf sex (0.04: -0.28, 0.35) also was not associated with serum Brix, indicating that physiological ability to absorb IgG does not differ by sex (Gomes et al., 2026; Sağlam et al., 2026). Consequently, as colostrum Brix substantially increased the probability of a calf achieving a serum Brix class higher than poor class (Fig. 2c), with most classes classified as either poor or excellent (Fig. 2d), corroborating to findings that passive immunity transfer is driven by colostrum management practices (*e.g.*, quality, volume, and timing) rather than calf size or gender (Lombard et al., 2020).

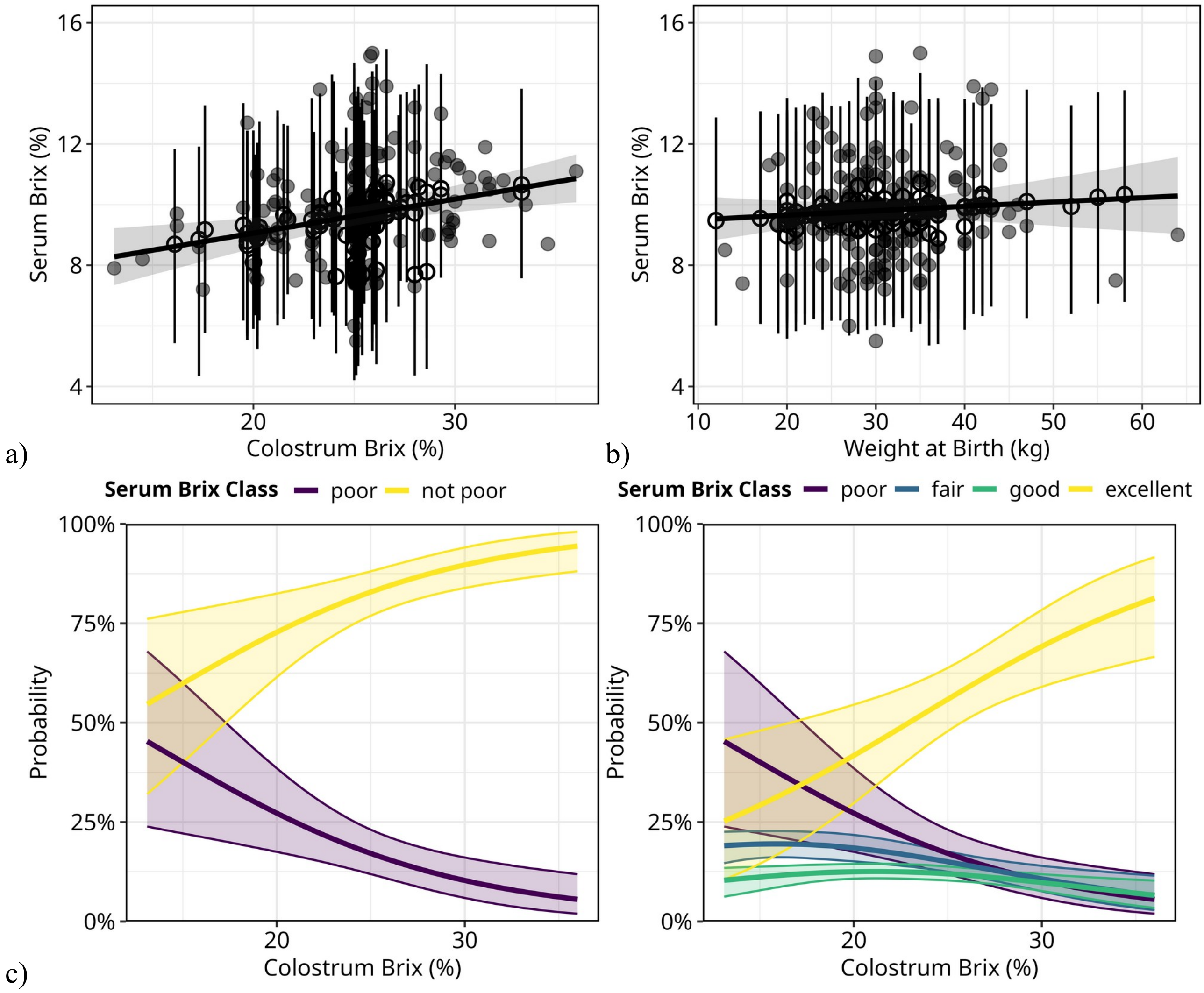

Figure 2. Posterior predictions show a linear relationship between calf serum Brix (%) and colostrum Brix (%, a) and a null relationship (*i.e.*, posterior mean included a horizontal line) with weight at birth (kg, b). As colostrum Brix increased, the probability of serum Brix belonging to a quality class higher than poor increased (c), with most classes classified as either poor or excellent (d). Solid lines: posterior mean; shaded areas: 95% credible intervals; filled points: measurements; open points: best imputed points; vertical lines: 95% credible intervals for the imputed points.

Similar to previous studies (Davies et al., 2025; Pereira et al., 2024), serum Brix quality class was positively associated with disease event rate (Fig. 3ab) and the cumulative risk of a disease event (Fig. 3c), showing an increased disease event rate for the poor class when compared with the excellent class (Fig. 3d), especially as animals get older (days > 50). Previous number of diseases (Fig. 4a; 0.17: 0.08, 0.25) was also associated with the disease event rate, suggesting that disease history may reflect persistent individual susceptibility, incomplete recovery, repeated exposure to shared risk factors, or impairment of immune competence caused by previous disease (Svensson et al., 2006; Lindley et al., 2025). Weight at birth (Fig. 4b; -0.23: -0.38, -0.09) was negatively associated with the disease event rate, indicating that heavier calves might have developed a stronger immune response (Moreira et al., 2024; Davies et al., 2025). Sex (-0.28: -0.57, 0.01) was not associated with disease event rate. Agrivoltaics installation (agrivoltaics pre-weaned: 0.07: -0.39, 0.52; agrivoltaics post-weaned: 0.58: -0.05, 1.20) were also not associated with the disease event rate, indicating that agrivoltaics can be implemented as a safe, dual-use system for increased calf comfort and welfare.

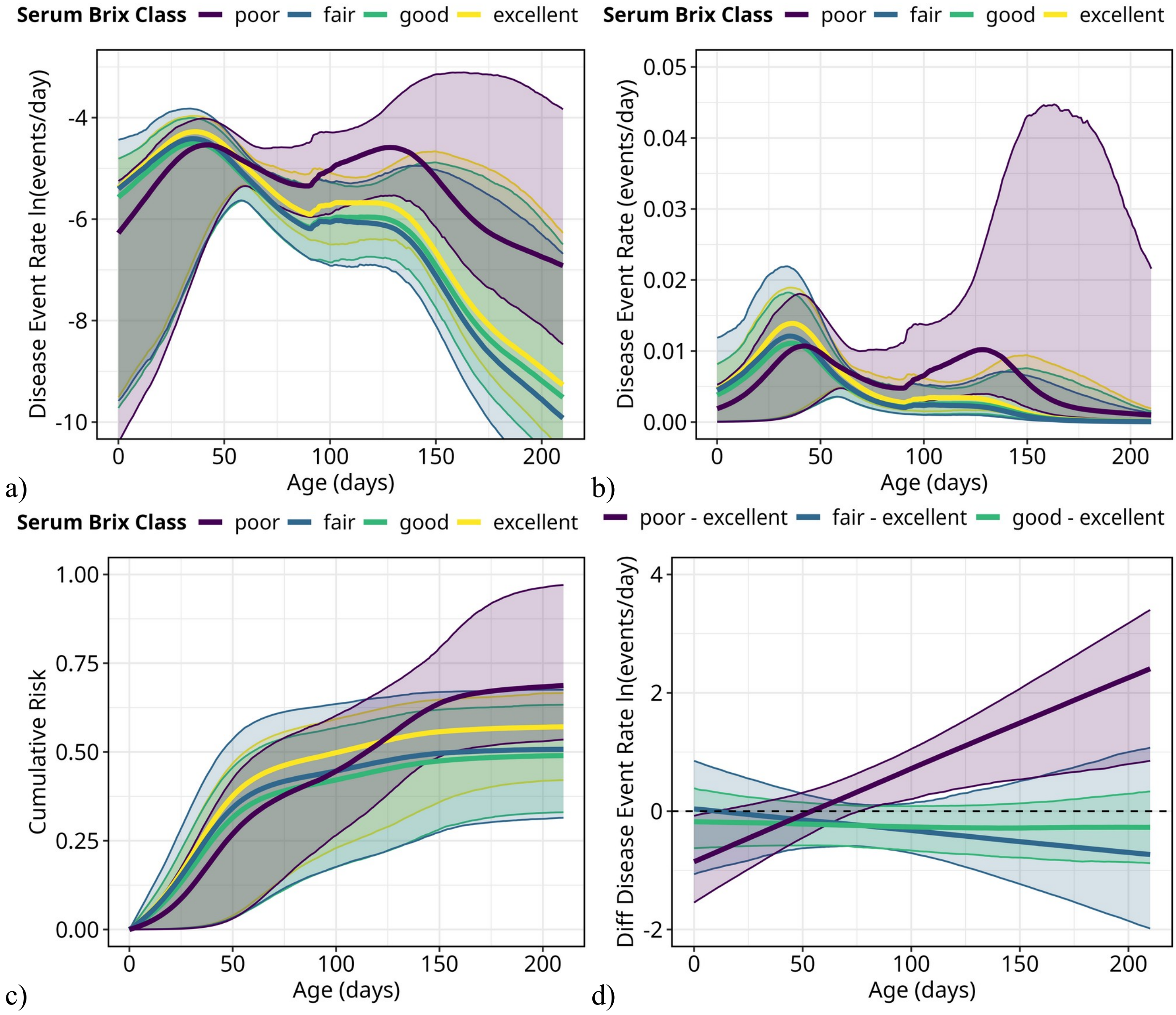

Figure 3. Daily disease event rates (link function, a; response scale; b), cumulative risk of disease event (c) and posterior contrasts (difference) of the link function (d) for serum Brix class (poor < 8.1%, 8.1% ≤ fair < 8.9%, 8.9% ≤ good < 9.4%, excellent ≥ 9.4%; Lombard et al., 2020) for calves raised in the agrivoltaic system indicate that calves with poor serum Brix exhibit substantially higher disease susceptibility. Solid lines: posterior mean; shaded areas: 95% credible intervals; dashed horizontal lines: not statistically different line.

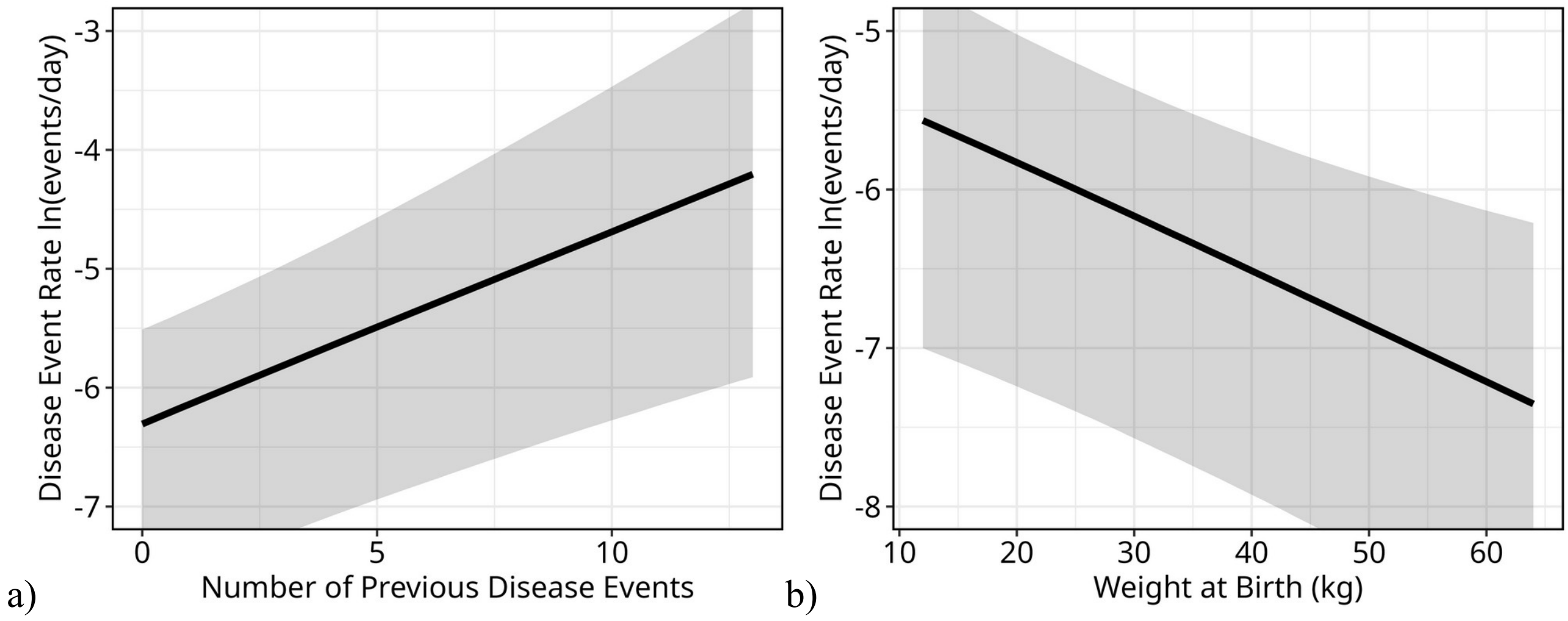

Figure 4. Number of previous disease events (a) increased the disease event rate, while weight at birth (b) decreased it. Solid lines: posterior mean; shaded areas: 95% credible intervals. Note that the shaded areas include the uncertainty for the intercept and the slope.

While male calves were heavier at birth ($\beta_{G3}$= 0.05: 0.01, 0.10 for males), serum Brix quality class showed no clear association with calf weight (Fig. 5ab; -0.18 to 0.16 for all $\beta_{G3}$ and $\beta_{K3}$), indicating that proper detection and disease treatment were sufficient to overcome the negative effects of low serum Brix quality.

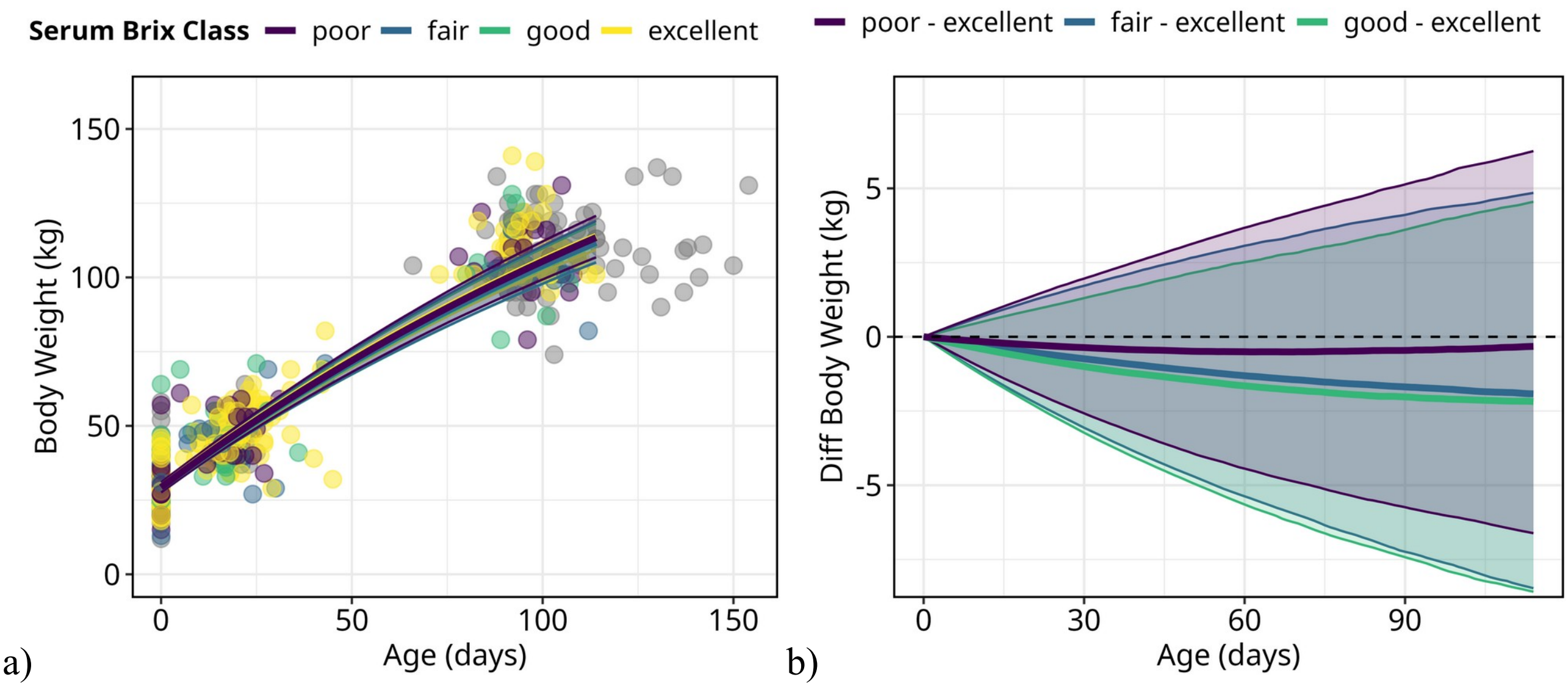

Figure 5. Body weight (kg) of calves did not show statistically substantial effects of serum Brix quality. Gray points: points with missing serum Brix; solid lines: posterior mean; shaded areas: 95% credible intervals.

## 4. Conclusions

We implemented a hierarchical Bayesian analysis using health and performance records of newborn dairy-calves from a commercial dairy farm. This approach demonstrates how advanced analytics can foster Livestock 5.0 in existing commercial operations by handling missing, real-world data. Through this statistical framework, we found that serum Brix was positively driven by colostrum Brix and was not influenced by weight at birth. The risk of disease substantially increased for calves with poor serum Brix quality class, especially as they grew older (> 50 days) calves, indicating that the detrimental effects of failed passive of immunity transfer persist beyond neonatal period. The disease risk were further compounded by the cumulative effects of previous diseases, whereas calves that were heavier at birth demonstrated a lower risk of disease.

Importantly, agrivoltaics installation did not increase the risk of disease, neither before nor after weaning, confirming its viability as a safe, dual-use system for increased calf comfort and welfare. Interestingly, serum Brix quality class did not affect animal weight, suggesting that detection and treatment of early disease signs might counteract animal performance loss due to poor immunity transfer. Nevertheless, because diseases increase farming costs through increased labor and medications, farm management should prioritize maximize colostrum Brix to reduce poor passive immunity transfer and, consequently, reduce disease event rates.

**Data availability statement**

Data and code (with all intermediary statistical analyses) are available online (Milan, 2026).

**Acknowledgments**

São Francisco Farm for providing the data. Funding: Brazilian National Council for Scientific and Technological Development (CNPq; Proc. 446340/2024-3, 314878/2025-4, and 442650/2025-6), and São Paulo Research Foundation (FAPESP; Proc. 2026/06064-0).

**Authors contributions**

Conceptualization: FAD, AAO, JFS. Methodology: AAO, HFMM. Software: TATS, HFMM, Formal Analysis: FAD, HFMM. Resources: FAD, AKS, ASCM, HFMM. Data Curatino: FAD, AAP. UB. Writing – Original Draft: FAD, HFMM. Writing – Review & Editing: all authors.

**Ethics Statement**

Authors declare no conflicts of interest.

**Declaration of Generative AI**

Statistical code in R and Stan were debugged and optimized with the support of Generative IA (ChatGPT, Gemini, Claud, and DeepSeek). All code was revised prior to execution and authors maintain full responsibility for the analysis.